**Title:** A stochastic simulation of the dislocation-mediated etching of porous GaN distributed Bragg reflectors

**Authors:** Piotr Sokolinski, Ben Thornley, Zetai Xu, Thom R. Harris-Lee, Menno J. Kappers, Rachel A. Oliver

**Affiliations:** Department of Materials Science, University of Cambridge, 27 Charles Babbage Road, Cambridge, CB3 0FS.

**Abstract**

Distributed Bragg reflectors (DBRs) can be fabricated by electrochemically etching nitride epitaxial structures consisting of alternating layers of highly *n*-type doped and non-intentionally doped (NID) GaN.  Threading dislocations (TDs) can be electrochemically etched into transport pipelines that can carry the etchant through the NID layers to access the doped material.  Experimentally this has been shown to involve a mechanism where the etching pathway may follow one TD into a doped layer and then propagate sideways through the doped layer to continue via a different TD.  Across multiple layers this process creates complex pore structures that have been described as "cascades". Here, we build a stochastic simulation for the DBR etching process that can reproduce some key features of the observed microstructures including the cascade morphology. By comparing the simulation output to samples etched at a range of voltages, we show that we can reproduce variations in experimental chronoamperometry data with applied bias by varying the probability of etching the doped layers within the simulation. The outputs of the resulting simulations replicate the experimentally observed cascade morphology.  At higher voltages, experimental data reveal a lower proportion of cascade features, a trend that is also replicated by the simulations for relevant probability values. Outputs of the simulations also correlate well with experimental chronoamperometry data for samples where – unlike in a DBR – the thicknesses of the doped layers vary through the epitaxial multilayer, suggesting that the probabilistic simulation can be applied to a range of structures to help understand the dislocation-mediated electrochemical etching process.

I. INTRODUCTION

Electrochemical porosification of gallium nitride can result in a wide range of nanoscale structures that exhibit properties unobtainable in conventional nitride semiconductors.  To date, the reduction in refractive index related to porosity has been the most widely explored and exploited property change, finding widespread application in distributed Bragg reflectors (DBRs).  DBRs consisting of multiple alternating layers of porous and non-porous GaN have been employed to enhance light extraction from light emitting diodes[1,2] and single photon sources[3], and to create optical confinement in resonant cavity photodiodes[4], and in both optically[5] and electrically[6] pumped vertical cavity surface emitting lasers.

The electrochemical process which enables these developments is conductivity-selective:  in a starting structure consisting of non-intentionally doped (NID) and *n*-doped layers, only the doped layers are expected to be porosified[7].  The NID layers are left notionally untouched since they do not support the flow of current necessary to drive the process.  Hence, in the process flow, it is necessary to consider how the etchant will access the buried doped layers in the doped/NID stack. This has frequently been achieved by creating deep trenches through the stack defined using lithography and reactive ion etching, so that the etchant accesses buried doped layers at the trench sidewalls[8]. In this case, the top surface of the DBR, usually an NID layer, is covered with (for example) SiO$_2$. Although this method yields highly reflective DBRs, the trench processing adds expense and complexity.

More recently, an alternative approach was discovered in which no deep trenches are used and the surface is not coated with a protective layer. As in the process employing deep trenches, the starting point is an alternating stack of doped and NID layers with an NID layer at the surface. However, in this case, the etchant accesses the doped layers through channels formed during the etching process at the sites of threading dislocations (TDs)[9], defects that are ubiquitously present in heteroepitaxial gallium nitride layers. This approach allows rather uniform etching of whole wafers without prior lithographic processing[10]. These wafers can then be used as pseudo-substrates for subsequent overgrowth of device epitaxy[1], so that the porous material is quite straightforwardly incorporated into standard nitride device process flows.

The TD-mediated mechanism has been thoroughly evidenced by using scanning transmission electron microscopy (STEM) to image the nanoscale pipeline etched through the NID layers at atomic resolution[9]. Such atomic resolution images allow a Burgers circuit to be drawn around the pipeline, with the closure failure of the circuit proving that the pipeline is, in effect, an open core TD. The original interpretation of the STEM data was that every TD formed a pipeline that proceeded through the entire doped stack, allowing the etchant to access each doped layer in turn so that a field of porosity associated with every TD formed at every doped layer[11]. This interpretation has since been referred to as the "kebab model", an analogy in which the TD is likened to the skewer in a shish kebab, with fields of porosity strung along it like foodstuffs along the kebab skewer[12].

However, experimental data has increasingly revealed deviations from this picture. First, Massabuau *et al.*[13] showed that towards the bottom of a multilayer stack the fields of porosity become larger suggesting that some TD pipelines stop partway down the DBR. Next, Ghosh *et al.*[14] demonstrated that even in the first porous layer, not all the TDs that are present are necessarily involved in etching. More recently, Thornley *et al.*[12] have shown using focused ion beam (FIB) tomography that in some cases the etching pathway initially propagates vertically down a TD, then laterally through a doped layer, intersecting a separate, unetched TD. A vertical etching pipeline then forms at that new TD, rather than (or sometimes in addition to) the original TD. As a result, TD pipelines may become active (providing a pipeline for the flow of etchant) or inactive (unetched and providing no pipeline) at multiple points down the DBR stack. In this case, the etchant pathway may be described as being more like a "cascade" in morphology than like a kebab. However, some structures that appear to correspond to the original kebab model are still observed in the same samples, alongside cascades. The different pore morphologies implied by the FIB tomography data are illustrated in Figure 1.

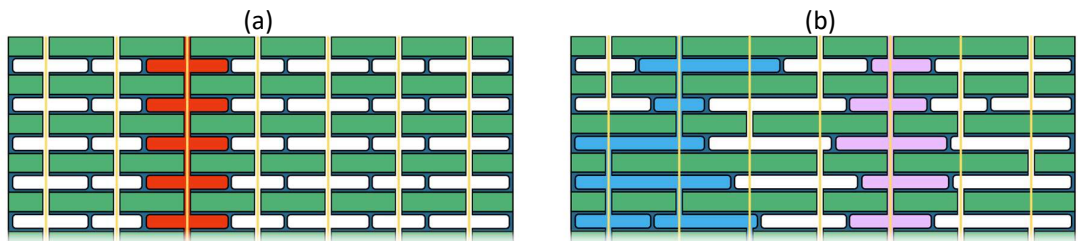

FIG. 1. Schematic illustrations of: (a) Porosification via TDs (marked in yellow) within a model where every TD forms a pipeline for the etchant to access every doped layer ("the kebab model"). The pore structure created for one of the individual TD pipelines initiating at the surface is marked in red. (b) Porosification via TDs within a model where not all TDs are active in bringing the etchant to every layer, but there is some possibility that a laterally propagating pore will intersect an inactive TD and that inactive TD will then become active to carry the etchant to a subsequent layer ("the cascade model"). The pore structure created for one of the individual TD pipelines initiating at the surface is marked in blue. Also marked (in purple) is a TD that, despite being in a sample where the cascades are in operation, still exhibits the kebab morphology.

Currently the physics of TD-based etching are not fully understood. However, we propose here that it is nonetheless possible to gain insights into the factors controlling the TD-mediated porosification mechanism by employing a stochastic simulation, where we assign separate values to the probabilities of etching doped layers and the region around the TD. In this manuscript, we develop such a simulation and compare its outputs to DBR samples etched at a range of voltages, and to other doped/NID multilayers with varying layer thicknesses in the stack. We also discuss the simulation's relevance to a broader range of samples and geometries.

## II. METHODS
### A. Description of simulation

The simulation was built in Python3, and Claude.ai[15] was exploited to integrate the "numba" library in Python to allow use of just-in-time compiling to increase efficiency and scalability. A version of the relevant code is publicly available[16]. The evolving DBR structure is modelled as an array. Pixels in that array may be assigned as etchant, NID GaN, Si-doped GaN or TD. Each pixel corresponds to a nominal area of 5 nm × 5 nm. At the onset of the simulation, the array (which is illustrated in Figure 3(a)) consists of alternating 50 nm layers of doped (blue in the Figure) and NID (green) GaN, with the topmost nitride layer being NID GaN. Above this is a layer of etchant (purple). TDs penetrate through the stack, and in the full simulation are modelled as vertical lines with spacing 500 nm (100 pixels) and width 5 nm (1 pixel). The latter value corresponds approximately to the width of a pipeline at a TD as observed by Massabuau *et al.*[9]. The pixels representing NID GaN are assigned a zero probability of etching. The pixels representing TDs and doped layers are assigned a finite probability of etching, where the probability of etching the doped layer ($P_{doped}$) is in general different to the probability of etching the TD ($P_{disloc}$) and can be changed between different complete runs of the simulation. Within the simulations described in this manuscript we maintain a constant value of $P_{disloc}$, but in principle that value could also be varied. We note that at positions where a TD crosses a doped GaN layer, the assigned probability is that of the doped GaN layer.

From the initial array described above, the simulation proceeded by searching for all pixels where etchant is present, and then searching the four nearest neighbours of those pixels, for an adjacent pixel where the probability of etching is greater than zero. Whether said pixels are etched at each iteration of the simulation is decided based on the probability assigned to that pixel. When a pixel is etched, it is replaced with an etchant pixel in the array. Repeated searches for etchable pixels are made, and the simulation continues until all etchable sites are exhausted. In order to represent the non-intersecting domains of porosity observed to be associated with each TD core in microscopy images of individual porous layers in real DBRs[9], we added an additional criterion to the code that prevents two neighbouring pores from merging to form one bigger pore.

Within the simulation, we assume that each pixel that is etched results in one (arbitrary) unit of charge being generated which leads to the flow of a current in the electrochemical etching circuit. This allows the simulation to be used to produce graphs of the expected variation in current flow as a function of simulation iteration (as an analogue for time), which can then be compared to experimental chronoamperometry data (see Section II.C). The simulated current-time plots presented in Section III were obtained by simulating different etching conditions for DBRs containing 300 TDs. Each set of simulated etching conditions was run three times, and the results averaged to give the overall model curve. The same simulations have also been used to derive populations of "kebab" and "cascade" features using the metrics developed by Thornley et al.[12] for analysis of FIB tomography data on real DBRs (as discussed further in section III.B). However, for illustrative purposes, smaller simulations have been used throughout to produce figures that exemplify features of our results, and instances of this are noted in corresponding figure captions.

### B. Epitaxial structures for comparator experiments

In places in this paper, we will draw comparisons with the samples previously described by both Ghosh et al.[14] and by Thornley et al.[12], who both worked on the same materials. However, for the chronoamperometry studies and some aspects of the structural characterisation, bespoke samples were grown for this study. All samples were grown by metalorganic vapour phase epitaxy (MOVPE) in an Aixtron close-coupled showerhead reactor on (111)-oriented 150 mm Si wafers using the methods described in Ghosh et al.[14].

Sample A consisted of the Si substrate, on which was grown a 250 nm AlN nucleation layer, a 1700 nm graded $Al_xGa_{1-x}N$ buffer (from x = 0.75 to x = 0.25), and a 670 nm NID GaN buffer. Next, five periods of a latent DBR structure were grown, each consisting of 63 nm of highly Si-doped (at 1 × $10^{19}$ cm$^{-3}$) GaN and 51 nm of NID GaN. We note that whilst this sample is rather similar to that used by Ghosh et al.[14], there are some differences in buffer structure and also the layers in the alternating stack of doped and undoped material are slightly thicker due to the samples having been designed to have different peak reflectance wavelengths.

Samples B and C used the same buffer structure as sample A. However, in these samples the thicknesses in the doped layers of the doped/NID GaN multilayer grown on the buffer vary. For sample B, the first doped layer to be grown has a thickness of 40 nm, and the doped layers increase in thickness by increments of 10 nm as the growth proceeds, so that the doped layer closest to the surface has a thickness of 80 nm. For sample C on the other hand, the first doped layer to be grown has thickness 80 nm, and the thickness of the doped layers decrease by increments of 10 nm as the growth proceeds, so that the doped layer nearest to the surface has a thickness of 40 nm. All the NID spacer layers sandwiched between doped layers, as well as the final NID cap, are 50 nm thick. The structures of all three samples are illustrated in Figure 2.

### C. Electrochemical Porosification and Chronoamperometry

All samples were divided into 1.3 cm × 2.5 cm chips for porosification, but no lithographic processing was performed. At one end of each chip, a deep scratch was made into the surface of the sample using a diamond scribe, penetrating through all 5 intentionally doped layers. An indium contact was then soldered onto the scratch, with the intention of providing electrical contact to all layers throughout the stack. To create the electrochemical circuit, the sample was partially submerged in a 0.25 mol/dm$^3$ oxalic acid solution, with the indium contact kept clear of the solution. A platinum counter electrode was used, and potential was applied using a Gamry 3000 potentiostat in a two-electrode configuration. The potential was held constant for individual etching experiments but varied between experiments and will be specified below. All applied potentials were between 6 V and 12 V. (We note that etching potentials used in this paper cannot be directly compared to those used by Ghosh et al.[14] since a different potentiostat was used in that case and slight voltage discrepancies may arise.)

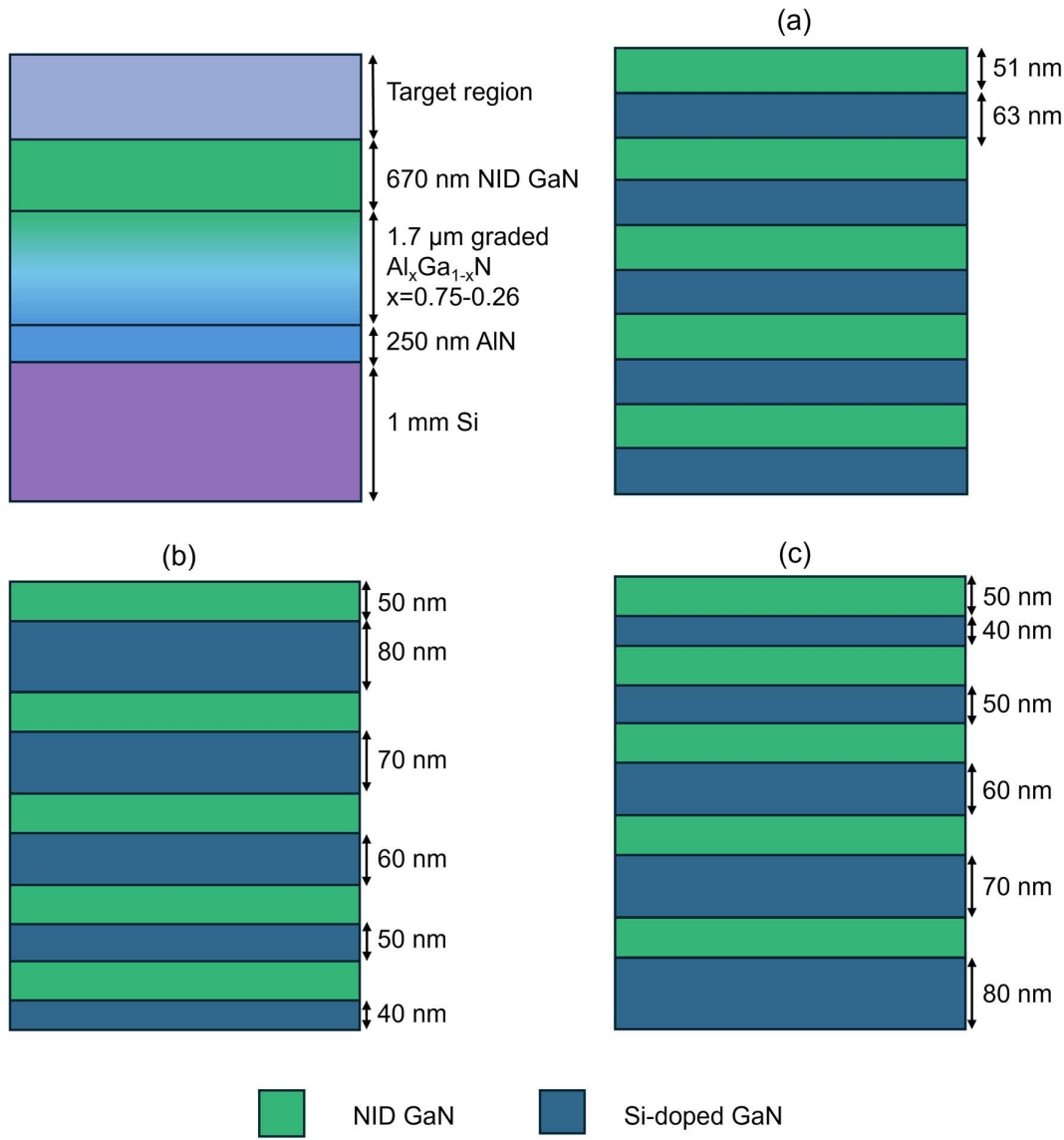

FIG. 2. Schematic diagrams of the wafers grown for this study. All wafers have the basic structure shown at the top left, with different target regions as follows: (a) all NID layers have thickness 51 nm and all doped regions have thickness 63 nm (Wafer A); (b) all NID layers have thickness 50 nm and the first doped layer to be grown has thickness 40 nm, with layer thicknesses increasing by 10 nm increments up the stack (Wafer B); (c) all NID layers have thickness 50 nm and the first doped layer to be grown has thickness 80 nm, with layer thicknesses decreasing by 10 nm increments up the stack (Wafer C).

The variation in current as a function of time was recorded throughout the etching process for each constant applied potential using the Gamry potentiostat – a technique known as chronoamperometry. The current values presented were directly measured without normalization to area, but all sample areas submerged in etchant solution were approximately 1.3 cm × 1.3 cm. For all experiments, the etching process was continued until the measured current significantly dropped relative to the etching current to a steady value (<75 µA).

### D. Structural characterisation

Pore morphologies following electrochemical etching were assessed using scanning electron microscopy (SEM) performed in a Zeiss Gemini SEM 300. Cross-sectional images were recorded on cleaved cross sections using the in-column secondary electron (SE) detector with an electron landing energy (LE) of 2 - 3.5 keV.

## III. RESULTS
### A. Example output of simulation

First, we will provide an example of the output of the simulation for a single set of etching probabilities. Figure 3 gives an illustration of how the simulation proceeds, using a smaller simulation consisting of 4 TD cores (each 1 pixel wide) separated by 100 pixels of GaN to highlight etching pathways and pore wall morphologies in the model. Figure 3(a) represents the unetched sample in the simulation. The NID layers are shown in green, the doped layers are shown in mid-blue and TDs are shown in yellow where they run between the doped layers. In this case, the probability of etching at the TD, $P_{disloc}$, was set at 0.05 (and this value is maintained throughout the simulations we will describe), whereas the probability of etching the doped layer, $P_{doped}$, is set at 0.6. Figure 3(b) shows the simulation after 300 iterations. By this stage, the doped layer closest to the surface has been entirely porosified, and pores have filled with simulated etchant (shown in dark purple). Figure 3(c) shows the simulation after the etch has proceeded further (for 550 iterations). The two layers closest to the surface have now completely porosified, the third layer from the surface is nearly fully porosified, but with some continuous mid-blue, unporosified regions remaining. One TD has run ahead of the others and provided an etch pathway to the fourth doped layer from the surface, where pores are starting to form. Figure 3(d) shows the simulation after etching has run to completion. The same simulation output is reproduced at a larger size in Figure 3(e) with specific etching pathways manually highlighted. In Figure 3(e), two etching pathways (and hence continuous pores) are highlighted. On the right, outlined in cyan, we see an etching pathway which conforms closely to the "kebab model", carrying etchant down through the first four layers beneath the surface and just into the fifth layer. The second highlighted etching pathway, in magenta, is more complex. Etching follows a pathway that initiates at the second TD from the left, but – for example – at the third layer from the surface, etchant propagating laterally undercuts the leftmost TD, so that no etchant that has followed the leftmost TD pathway reaches the fourth layer from the surface. However, where the etchant that has followed the pathway at the TD second from left reaches the leftmost TD, it restarts etching via that leftmost TD pathway. This behaviour, whereby a TD is deactivated as an etchant pathway to a layer and then reactivated at a layer deeper beneath the surface, is characteristic of what we describe as "cascade" behaviour. Overall, the simulation reproduces the experimental observation[12], that some TDs result in a field of porosity at each doped layer, whilst others are active in some layers but "switch off" and do not result in a field of porosity in other layers.

Another phenomenon is possible in the simulations, which we note here for completeness although it is not fully illustrated by Figure 3. An etchant pathway propagating laterally can intersect a currently inactive TD and under some circumstances create an etchant pathway that travels from a deeply buried doped layer to a doped layer closer to the surface (i.e. the etchant propagated vertically *upwards* in the figure), achieving etching in that doped layer if it has not locally yet been reached by another etchant pathway. This is occasionally seen in our simulations, as a consequence of the assumptions of our code, but we have not yet found evidence of this phenomenon in experimental data. This point will be relevant to a small element of our later discussion.

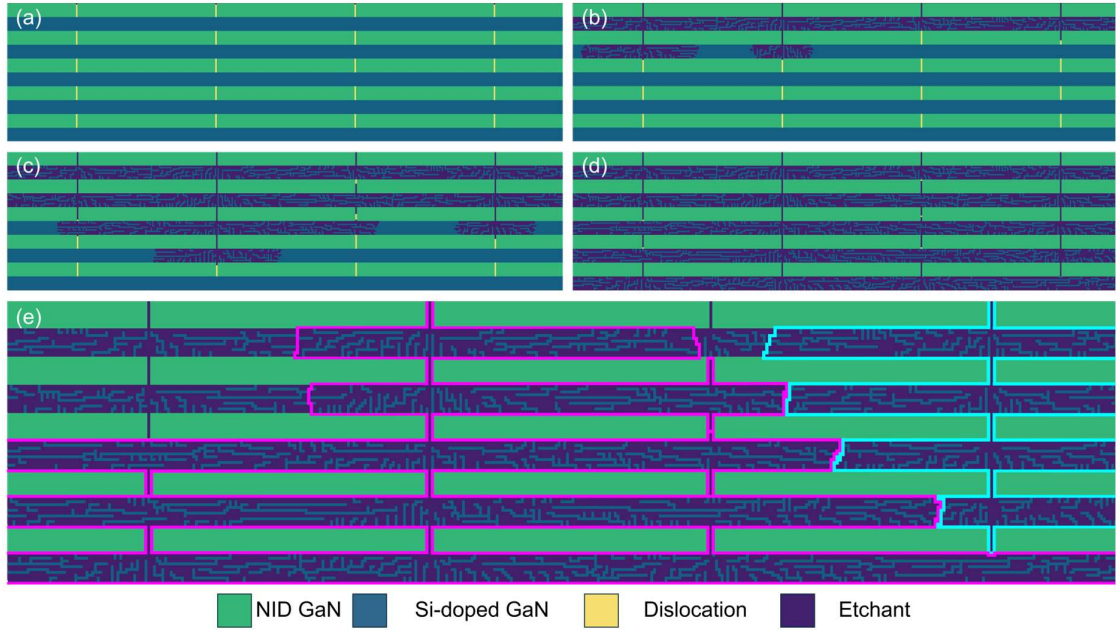

FIG. 3. Simulation outputs for an illustrative simulation of reduced size at different time points in the electrochemical etching process: (a) Prior to etching, (b) after 300 iterations of the model, (c) after 550 iterations of the model, (d) once no further etchable sites can be found. (e) is a larger version of output shown in (d) in which two continuous porous regions are highlighted, with the magenta lines highlighting a continuous pore formed in a cascade which initiates at one TD but penetrates laterally to two other TDs, and the cyan lines highlighting a kebab structure.

Figure 4 shows an output of the full simulation using 900 TDs in total. It is a simulated chronoamperometry curve, again for $P_{disloc}$ = 0.05 and $P_{doped}$ = 0.6. For this set of probabilities, we see an initial period during which little or no current flows, after which the current rises to a peak and then drops again, but does not return to zero. The current then oscillates, so that 5 peaks in current are seen, but the peaks become smaller and less distinct as the etch proceeds. The 5 peaks appear to correspond to the 5 layers of the DBR. Indeed, Massabuau et al.[9] reported layer by layer etching, for a 10 layer porous GaN DBR on a sapphire substrate, formed via the TD mediated method. In their studies, sample cross sections were created and imaged after etches of different durations, so that an apparent signature of layer-by-layer etching was observed in cross-sectional SEM.

These initial results suggest that the simple probabilistic simulation reproduces some key features of literature data, including the existence of both kebab and cascade structures. We note that Thornley et al.[12] saw that the prevalence of kebab and cascade structures depended on the etching voltage and suggested that this was due to the rates of dislocation and doped-layer etching changing relative to one another when voltage was varied. Hence, in the next section, we will discuss a series of experiments in which sample A was etched at a range of different voltages, and we will compare the resulting chronoamperometry data to the outputs of simulations with a range of different etching probabilities. In addition to drawing comparisons to etching, experiments performed on sample A, we will also draw comparisons with a voltage series reported by Thornley et al.[12] on similar samples.

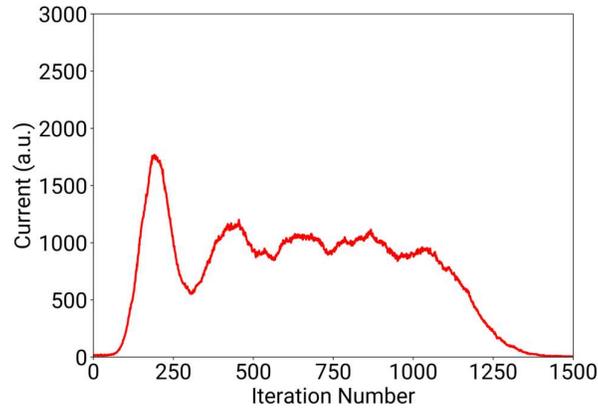

FIG. 4. Simulated chronoamperometry curve for Sample A for $P_{disloc}$ = 0.05 and $P_{doped}$ = 0.6.

### B. Comparing experiment and simulation for Sample A etched at different voltages

Here, we will first introduce the microstructures observed following etching of Sample A at different voltages and then introduce experimental chronoamperometry curves for those same voltages, which will be compared to relevant outputs from the model, using different values of $P_{doped}$. Sample A was etched at a range of voltages: 6 V, 7 V, 8 V and 12 V. SE images of cleaved cross sections showing the resulting porous microstructures are displayed in Figure 5. Even though the etching process was allowed to continue until the current significantly dropped relative to the etching current to a steady value (<75 µA), the samples have not necessarily etched completely. Whilst all samples show four porous layers, the fifth layer has not etched to the same extent as the other layers. Samples etched at 6 V and 7 V (Figure 5(a) and (b)) exhibit small pores whose lateral dimensions are often less than the width of the porous layer. As the etch voltage increases (to 8 V, Figure 5(c)) some pores of larger lateral extent are observed, whilst for etching at 12 V (Figure 5(d)) pores extend laterally for several hundred nanometers and in places NID layers have collapsed because of the large cavities beneath them. Also, at 12 V we observe pronounced etching of the NID layers, which is less prevalent (although still occasionally present) at lower etching voltages). These results are consistent with those previously reported by Ghosh *et al.*[14].

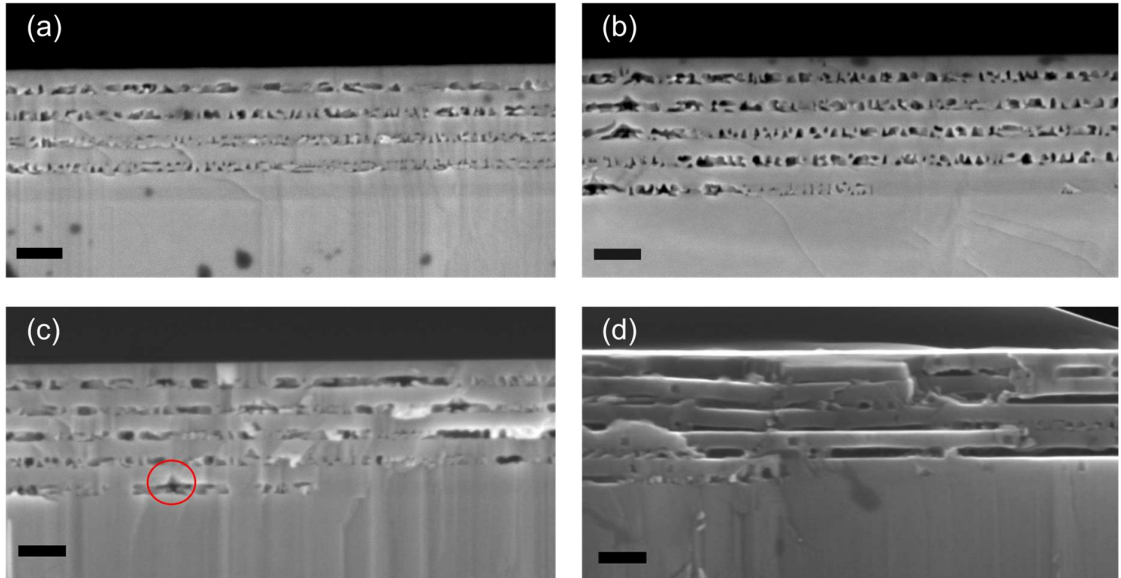

FIG. 5. Cross-sectional secondary electron images of sample A (DBR structure) etched at (a) 6 V, (b) 7 V, (c) 8 V and (d) 12 V. The scalebars in each of the images correspond to a length of 200 nm. Circled in red is conical pore opening up in the NID layer just above the doped layer, at the TD site as observed by Thornley et al.[12].

Figure 6(a) presents chronoamperometry data for etching at the same 4 voltages. For all the curves, we see an initially high current, which rapidly drops within the first ~200 s of the etching time. For the sample etched at 6V, the current remains fairly low throughout the etching process, after this initial spike. For the sample etched at 7 V, the current rises gently to a low peak (of about 0.43 mA) at around 485 seconds and then decreases gradually thereafter. As the voltage is increased to 8 V, we see that oscillations appear in the data. An initial peak arises at ca. 350 s, and then four smaller peaks appear fairly periodically thereafter, with a period of about 300 s. The peak heights gradually reduce as time proceeds. These data rather resemble the simulation output in Figure 4, except for the fact that the model does not show the initial current seen in the experiment. For the 12 V sample, oscillations are again observed, although the average current is higher and the five peaks are all more pronounced.

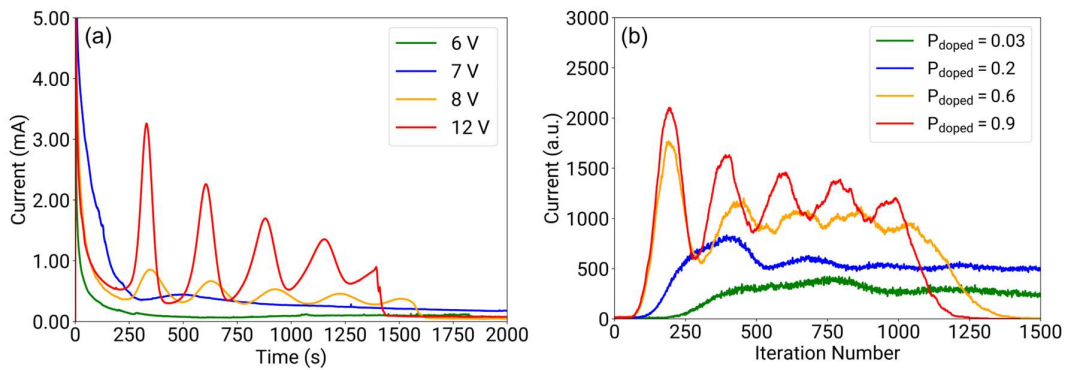

FIG. 6. (a) Experimental chronoamperometry data for Sample A etched at voltages between 6 and 12 V. (b) Simulated chronoamperometry data, for which $P_{disloc}$ = 0.05 in all cases and $P_{doped}$ varies from 0.03 to 0.9.

We have explored the initial high current that is seen in the experiment but not the simulations carefully, and note that it can be eliminated by coating the back and sidewalls of the sample with a polymer layer that the etchant cannot penetrate. In this case, where etching only occurs through the top surface of the sample (the situation mimicked by the simulation) the current in the first ca. 200 s is negligible. Data supporting this point are shown in the Supplementary Information in Figure S1).

Having thus eliminated a major discrepancy between the simulation output in Figure 4 (for which $P_{disloc}$ = 0.05 and $P_{doped}$ = 0.6) and the chronoamperometry curve for etching at 8 V, we next explore the impact of changing the etching probabilities in the simulation, and its impact on the simulated chronoamperometry curves. We have explored a range of probabilities and have selected parameters that provide a reasonable match to the variations seen in the data. The output of this exploration is given in Figure 6(b). Here we have decreased $P_{doped}$ whilst maintaining $P_{disloc}$ constant, to achieve results that to some extent mirror the impact of decreasing the etching voltage from 12 V to 6 V. Before discussing these simulated curves further, we stress that we are not claiming here that these probability values represent an accurate likelihood for the etching of any particular volume, merely that by varying the probabilities in the simulation it is possible to cause variations in the simulation output similar to the variations caused by varying the etching voltage in the experimental situation. We also note that the TD spacing in the simulation is another parameter that affects its output. Other combinations of probabilities and TD spacings may lead to similar outputs to those shown here.

Within these caveats we see that as we decrease $P_{doped}$ from 0.9 to 0.5 we see a slight reduction in the simulated etching current and achieve a reduction in the peak-to-trough current change in the oscillations. This is qualitatively similar to the change observed when decreasing the etching voltage from 12 to 8 V. A decrease in $P_{doped}$ to a value of 0.2 decreases the current further and causes the oscillations to die away to the point where they are barely identifiable beyond the first peak. As in the data, we see that the position of this first peak is now delayed relative to the first peak seen at the probabilities that correspond to higher etching voltages. Further decrease in $P_{doped}$ (to a value of 0.03) causes the current to die away still further, giving a somewhat analogous decrease in current to that seen when we reduce the voltage the lowest value examined here. Overall, reductions in $P_{doped}$ yield similar trends in the shape of the simulated chronoamperometry curves to the impact of reducing the applied bias in the real chronoamperometry curves, so that Figures 6(a) and 6(b) rather resemble one another.

In both the experimental data and the simulated curves, we see that for the highest bias (or equivalently high $P_{doped}$) we maintain fairly clear current oscillations throughout the etching process, although the peaks heights do drop as etching proceeds. For the next highest values of bias/$P_{doped}$, we see a more pronounced dying away of the oscillations. Our further inspection of the output of the simulation suggests that this is because, across the hundreds of TDs modelled, the etch front does not move uniformly into the sample. Instead, the etching of some TDs reaches deeper layers ahead of other TDs, as may be seen in Figure 3. Hence, the modelled etch current does not drop to a near zero value between peaks as there is never a moment when no doped layer is undergoing etching. Also, because some TD pipelines run ahead of others, when other TD pipelines reach these lower lying layers, there is less etchable material remaining, and the maxima for these layers thus do not reach such high current values. Hence, although at higher etching voltages / higher $P_{doped}$ etching may occur broadly layer by layer (as shown by Massabuau *et al.*[9]), the current peaks that may to some extent be assigned to individual layers actually involve the superposition of currents arising from multiple layers. At the lowest values of $P_{doped}$ (corresponding to the lowest applied voltages), lateral etching happens at a similar rate to vertical propagation of the etchant down the TDs so that etching can no longer be described as layer-by-layer and very little oscillation is seen.

Having shown that the simulation can (via variations in the probability of etching the doped layer) qualitatively reproduce the trends seen in the chronoamperometry data for varying etching voltages, we now assess whether key features of the microstructures can also be reproduced. Here, in order to access more detail of the microstructure than can be seen in the cross-sectional SEM images in Figure 5, we refer to the FIB tomography data sets recorded by Thornley *et al.*[12], which also address samples etched at a range of different voltages, although (as noted above) the samples have some subtle differences, and etching voltages may not be exactly comparable. To help with this comparison, we will first introduce metrics that allow some quantification of the extent to which the microstructure is made up of "kebabs" or "cascades".

If a TD acts as an etchant pathway to all layers, this is considered to conform to the "kebab" model. We refer to such TDs as "K5" TDs since they create a "kebab skewer" through all five porous layers. If a TD is active and is providing an etchant pathway at one or more porous layers, then does not provide an etchant pathway through the NID layer to a subsequent porous layer, but does provide a pathway to another layer further down the stack, this must mean that cascade behaviour is observed, and we define such a TD as a "C". Referring to Figure 3, the leftmost TD acts as a pathway from the surface to the first porous layer and the second porous layer but does not take etchant from the second porous layer to the third. However, the third porous layer in the region of this TD is etched, and for this to be true another TD must have been involved, in this case the TD second from the left. The leftmost TD then becomes active again and carries etchant to the fourth porous layer. For this to occur, we see that the etchant had to travel laterally through the third porous layer to restart etching down the leftmost TD pipeline from the third layer to the fourth, through the intervening NID layer. The deactivation and later reactivation of the leftmost TD is hence a cascade signature. This signature of cascade behaviour can be spotted in the simulation as a termination of the etchant pathway within a TD where it crosses between 2 doped layers through an NID layer (i.e. 1 TD pixel remains unetched in the NID layer). This unetched pixel is then counted automatically by our code. In Thornley *et al.*[12]'s data, such features were counted manually. We note that the sum of K5 TDs and C TDs is not necessarily 100%, since there is also the possibility of K1, K2, K3 and K4 TDs. Considering the general case of a "KX" TD, where X is an integer from 1 – 5 inclusive, such a TD would form porous fields in X adjacent layers, but would not form any porous fields in any non-adjacent layers, such that the etching pathway at that TD is continuous. The sum of all the KX TDs and the C TDs, would add up to the total number of active TDs in the sample. Examples of various KX TDs, and a C TD are shown in Figure 7.

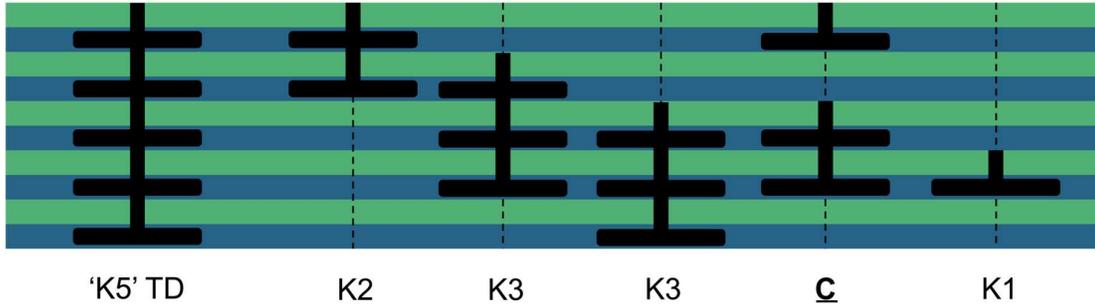

'K5' TD      K2      K3      K3      **C**      K1

Fig. 7: An illustration of KX and C TDs, showing in each region only etching facilitated by the closest local TD. Etchant is shown in black, doped layers are shown in mid blue, and NID layers are green. TDs that are not locally acting as etchant pipelines are shown as thin dashed lines. A 'K5' TD acts as an etchant pipeline through all doped layers, whereas other KX TD only carry etchant to X adjacent layers. For all KX other than K5, more than one possible configuration exists. For example, we show two different possible K3 TDs here. A C TD is active in a layer or layers near the surface, becomes inactive, so that it doesn't carry etchant to one or more doped layers, and then becomes active again further down the stack.

Thornley et al.[12] showed that as the applied voltage increased from 5 to 10 V, the percentage of all TDs that followed the K5 pattern increased and the percentage that followed the C pattern decreased. This is shown in Figure 8(a). Figure 8(b) displays an equivalent assessment of the frequency of K5 and C features in the simulation as a function of $P_{doped}$. For $P_{doped}$ between 0.05 and 0.5, little variation is seen, but as $P_{doped}$ increases from 0.5 to 0.9, we see the proportion of C features decrease and the proportion of K5 features increase. This is similar to the experimental observation with increasing voltage, and is consistent with our earlier suggestion (based on comparison of the simulated and experimental chronoamperometry data) that higher values of $P_{doped}$ within the simulation correspond to higher etching voltages.

Our simulation, as presented so far, is successful in that it predicts the formation of both kebab and cascade structures in the TD-mediated electrochemical etching of porous GaN DBRs, and can provide a reasonable match to the chronamperometry data arising from such etching experiments at a range of voltages. Based on the chronoamperometry data, an increase in the applied voltage in an experiment is simulated by an increase in the probability of etching the doped layer. Moreover, the same trend in probabilities derived from the comparison with the chronoamperometry data also yields trends in microstructural metrics that somewhat resemble those revealed by focused ion beam tomography characterisation of DBRs etched at varying voltages.

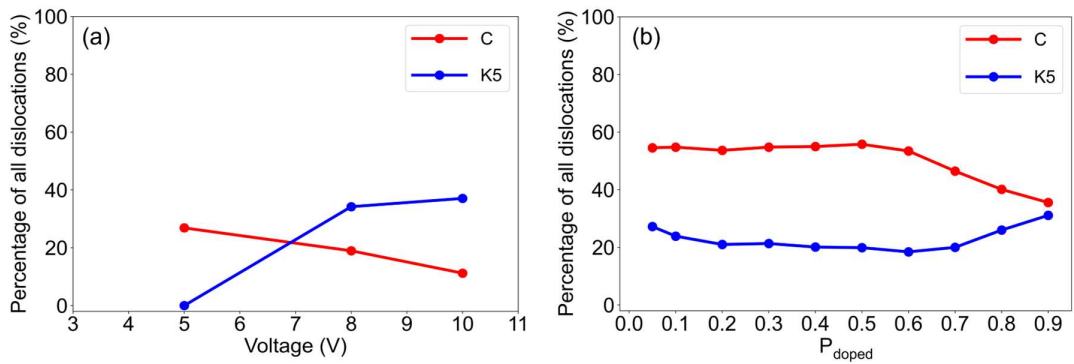

FIG. 8. (a) The percentage of all active TDs in the dataset of Thornley et al.[12] that form either K5 or C structures, as a function of etching voltage. (b) The percentage of all active TDs in our simulation that form either K5 or C structures as a function of $P_{doped}$.

However, the simulation also has several shortcomings revealed by comparison to the datasets in this section. The simulation does not predict the failure, particularly at low voltages, to completely etch the lowest lying layer beneath the surface. (However, we note here, that in Figure 6(b), the simulated current has not dropped to zero at the longest times displayed, although it would, if longer timescales were displayed. The simulation suggests etching at low voltages is a very slow process yielding low currents, and it is hence possible, that in the equivalent experiments, we did not etch the samples to completion). The simulation also does not predict the changes in pore morphology that are observed upon increasing the etching voltage. In the simulation, for all values of $P_{doped}$, we see finely structured tortuous pores, whereas in the experimental data set the pores are small with fine structure at low voltages, but become large and open at higher voltages (see Figure 5). A further failure to predict the microstructure is seen in the shape of the pores where the TD pipeline penetrates into the doped layer. Thornley *et al.*[12] observed experimentally that, particularly at high voltages, a conical pore opens up in the NID layer just above the doped layer, at the TD site. An instance of this phenomenon is highlighted in Figure 5, by a red ring. However, away from the TD line, the simulation currently constrains the probability of etching the NID layer to zero and there is thus no way it could reproduce this feature.

The strengths and weaknesses of the simulation described above suggest that it can provide some specialist (but incomplete) microstructural insights, and also that it can predict the approximate form of chronoamperometry curves to an extent which could aid their real world interpretation. This motivates the work in the next section to address whether the simulation can provide helpful insights into the chronoamperometry data for the electrochemical etching of samples beyond the simple DBR structure.

### C. Comparing experiment and simulation for samples with different layer thicknesses

The previous section discussed the comparison of real chronoamperometry data for a DBR sample (sample A) with relevant simulation results. This section will address the samples with varied layer thicknesses (samples B and C, see Figure 2(b) and 2(c))) again comparing experimental chronoamperometry data and simulation results. Figures 9(a) and 9(b) show experimental chronoamperometry data for samples B and C, respectively, for comparison with Figures 9(c) and 9(d) which show simulated chronoamperometry traces for equivalent structures at $P_{doped}$ = 0.7. (This value of $P_{doped}$ has been shown to roughly correspond to 10 V in studies of the DBR sample (Sample A)).

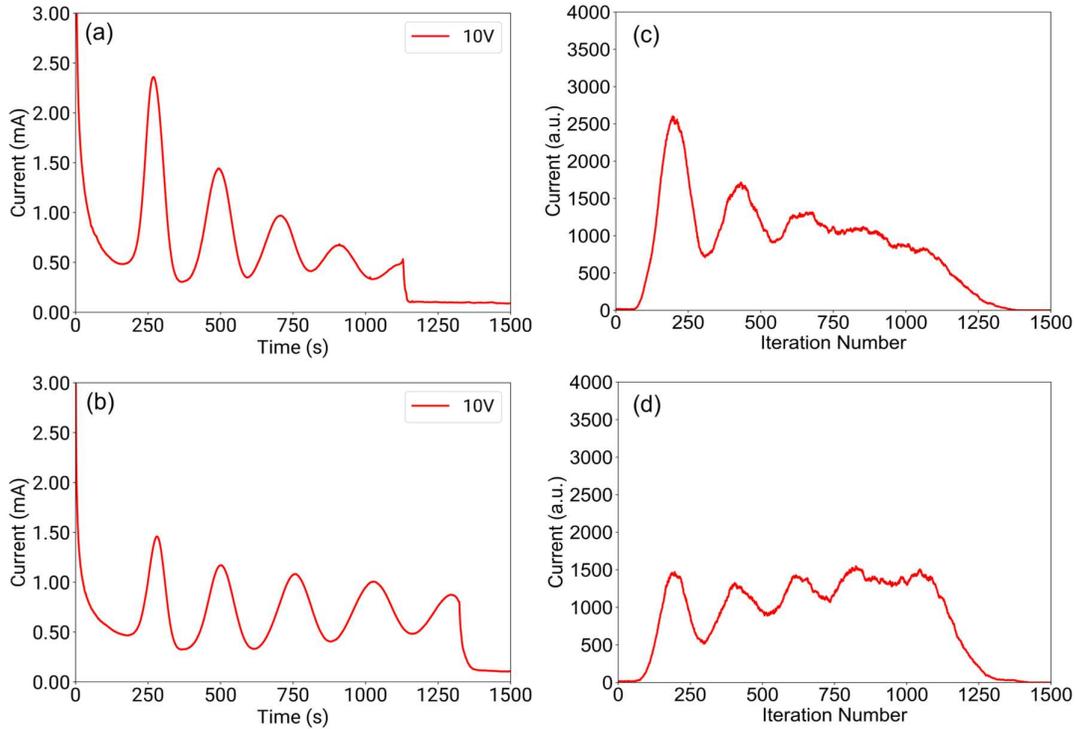

FIG. 9. On the left - experimental chronoamperometry data for the etching of (a) Sample B, and (b) Sample C; on the right simulated chronoamperometry traces for the etching of (c) Sample B and (d) Sample C. The reader is reminded that for sample B the doped layers are thicker near the surface of the sample, whereas for sample C the doped layers are thinner near the surface of the sample. For both simulations $P_{disloc}$ = 0.05 and $P_{doped}$ = 0.7.

In the experimental data, we again see the initial high current at very short times, which we have previously shown (see section III B) can be eliminated by preventing the etchant from accessing the back and sides of the sample.  Considering sample B, where the first doped layer to be grown is thinnest and the doped layer nearest the sample surface is thickest, both the experimental data and the equivalent simulation show oscillations whose amplitude decreases with time as we etch further into the stack.  Considering sample C, where the doped layer nearest to the surface is thinnest, the experimental data shows a fairly strong initial oscillation, but thereafter the peaks have fairly similar heights.  The slight decreases in peak height with etching time are much smaller than have been observed for other samples.  The simulation does not show such a pronounced initial oscillation, although the first peak in the series is higher and more pronounced than the next peak.  Thereafter, whilst peak heights remain similar as the simulation etch proceeds, a slight increase in peak height can be observed which is not present in the experimental data.

The fact that the simulation is successful in predicting the form of chronoamperometry data in a case where the etching current peaks are expected to decrease as the etch proceeds, but is less successful in the case where the etching current peaks are expected to increase as the etch proceeds, correlates with a shortcoming of the simulation that we noted above:  experimentally, lower lying layers etch less successfully than layers nearer the surface, but the simulation does not predict this. The less complete etching of lower lying layers will lead to lower peak currents in chronoamperometry, and this issue is highlighted more strongly in the case where the peak current is predicted by the simulation to increase slightly as the etch proceeds.

We also note that for both the experimental and the simulated chronoamperometry curves the variations in peak height should not be assumed to be solely related to the thicknesses of the etched layers. As noted in Section III.B, the etch front does not move uniformly through the sample. Hence, the current peaks that are associated with the thick sub-surface layers buried far beneath the surface in sample B will have a reduced amplitude, because parts of the relevant layers will already have been etched away by etchant that has arrived via TD pipelines that have (purely by chance) proceeded more quickly through the sample.

Overall, the simulation reproduces some of the differences between etching samples with uniform layer thicknesses and those with non-uniform layer thicknesses, and our discussion illustrates that deviations of the experiment from the simulation may help identify shortcomings in the etching process.

## IV. DISCUSSION

Thus far, the values of $P_{doped}$ and $P_{disloc}$ that we have employed are essentially fitting parameters that allow us to achieve a reasonable match to the experimental etching data. Next, we will briefly discuss the physics of etching and how variations in these probabilities with voltage might arise. Electrochemical etching of gallium nitride in the porous regime has been suggested to require the Zener tunnelling of holes through a space charge region in order to provide a supply of holes for anodic etching[17]. The probability of Zener tunneling increases as the local field increases, so that the suggestion arising from the comparison of our simulation and our experimental data – that the probability of etching the doped layer increases as the applied voltage increases – makes physical sense. Understanding the probability of etching a TD in an NID layer is more challenging. Where a TD intersects the sample surface, a small pit will form[18,19] and this pit can act to concentrate the local field, increasing the probability of etching. However, Zener tunneling requires not only a high local field, but also a high dopant density to allow holes to tunnel into the conduction band through the space charge region. If the NID spacer layers in our samples were entirely free from dopants, this would imply that tunnelling must occur through the entire NID spacer layer, which is rather unlikely for spacer layer thicknesses of 50 nm or more. However, there may be unintentional dopants in the NID layer (for example oxygen), and these dopants might segregate to a region adjacent to the TD core, and help mediate Zener tunneling. It is also possible that states in the bandgap associated with the TD core mediate trap-assisted tunnelling or another alternative current flow mechanism. Further insight into the detailed mechanism of the etching out of the nanopipe at the TD would be required to understand the actual probabilities of etching and how they vary with voltage (and indeed with other factors such as TD type). However, the foregoing discussion suggests that it is very reasonable to suggest that – as in the simulation – the probability of etching the TD region in the NID layer is lower than the probability of etching a doped layer under most etching conditions. We can also briefly consider what controls the conductivity where a TD intersects a doped region. TD conductivity levels related to either unintentional doping or states in the bandgap are expected to be rather low compared to the conductivities achieved via silicon incorporation in the doped layers. This justifies the fact that in our model the probability of etching at the TD in the doped layer is set as equal to that of the doped layer since the contributions to conductivity from the dopant will dominate.

Given that the stochastic simulation has had some success in reproducing experimental data for TD-mediated etching of DBRs, it is also interesting to speculate as to what insights it might give in other situations where electrochemical etching has been applied. For example, the simulation can be adapted to the situation where deep trenches cutting through the stack are defined using lithography and reactive ion etching, so that the etchant accesses buried doped layers from the trench sidewalls[8]. Figure 10 shows the output of an adapted version of the simulation, where the as-grown surface of the GaN (at the top of the figure) is blocked from the etchant, but an exposed cross-section is created cutting through the layers (at the left-hand side of the figure, highlighted in red).  Etching proceeds from the left-hand side, and separate etchant pathways have individual colours (dark blue, mid blue, teal green, lime green and yellow).  As in the original simulation, TDs run through the stack of doped layers.  The simulation has been run with $P_{doped}$ = 0.3 and $P_{disloc}$ = 0.05.  The TD spacing is 100 pixels (500 nm), as in the original simulation, but in the figure, the simulation has been compressed in the direction parallel to the layers to allow a significant lateral extent to be visible, so that the TDs appear as rather closely spaced faint pale lines.  For the first approximately 2850 pixels (14.25 μm) as the etch proceeds from the exposed cross section at the left, all porosification occurs along the doped layers with no involvement of TDs. However, as the etchant penetrates further laterally into the stack, a TD from the second layer beneath the surface (mid blue pathway) is etched through to the layer beneath (third layer from the surface) so that the teal green pore is blocked from making further progress.  Similarly, another TD pathway is created from the second layer to the doped layer above, so that mid blue pores dominate the top right-hand quadrant of the image.  TDs from the fourth layer from the surface (lime green pore) also penetrate to other layers, so that (for example) the yellow pore (furthest below the surface) is blocked from proceeding. Overall, at some distance from the exposed cross section, we see the onset of TD etching, given that there is some probability of opening up an etch pathway at the TD, and cascade behaviour then arises, even in this exposed sidewall geometry.

Focussed ion beam tomography suggests that the three-dimensional morphology of the pores close to such an exposed cross-section is quite different to that which is seen in the TD-mediated etching case[11].  Close to the exposed cross section, pores run roughly perpendicular to the cross section, and may extend over microns without deviating substantially in orientation.  However, the TD-mediated etching mechanism, under the same etching conditions, leads to pores of similar dimension that radiate from the relevant TD pipeline like the petals of a flower[11,12].  Griffin et al.[11] showed that the aligned pores seen close to the exposed cross section, render the porous GaN birefringent.

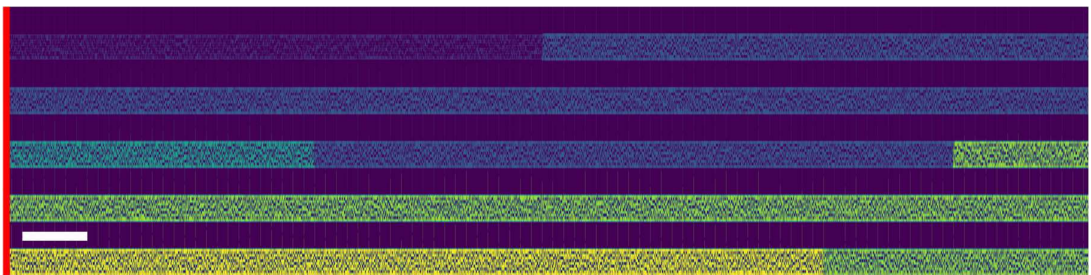

FIG. 10.  Output of a simulation for a situation where the as-grown surface of the GaN structure is blocked from contact with the etchant, but an exposed cross section is created cutting through the layers at the left-hand side of a figure, highlighted in red.  Separate etchant pathways have been given separate colours (dark blue, mid blue, teal green, lime green and yellow).  Note that, in order to fit on the page, the figure has been compressed by a factor of 25 in the direction parallel to the layers. The white scale bar at the bottom left corresponds to 3 μm in the lateral direction only.

 However, in their samples etched regions were observed, further from the exposed cross-section that did not exhibit birefringence, suggesting that the pores were no longer well aligned.  The simulation in Figure 10 may provide an explanation for this:  if TD pipelines are increasingly likely to be activated as larger distances from the exposed cross section, this will result in decreased alignment of the pores overall and particularly in pores in different layers not having the same alignment as pores immediately above or below them, leading to a lack of overall observed birefringence. Overall, this brief foray into application of our simulation to alternative etching geometries suggests that TD mediated etching may also occur in situations where it had not previously been considered relevant and may influence the properties of the resulting materials and devices in device-relevant ways.

## V. CONCLUSIONS

We have shown that a stochastic simulation can reproduce key features of both the chronoamperometry and the microstructural data arising from the etching of GaN DBRs via TD pathways. For example, where oscillations are observed in the chronoamperometry data, which die away as the etch proceeds, a similar decrease in oscillation amplitude was reproduced by the simulation. The simulation was then interrogated to reveal that this corresponds to a deviation from layer-by-layer etching when some TD pipelines run ahead of the main etch front and start to etch deeper lying layers. By manipulating the probability of etching the doped layers, in a manner consistent with the expected changes as the applied bias varies, we can change the extent to which etching proceeds layer-by-layer. These change to the doped layer etching probability also alters the fractions of "cascade" and "kebab" structures, consistent with the observations of previous detailed focused ion beam tomography studies[12]. Whilst the physics of TD etching remains unclear, we note that our success in simulating the etching at different voltages by varying only the probability of etching of the doped layer, suggests that the impact of applied voltage on doped layer etching and TD etching may not be the same, hinting at differences in the mechanism of these two parts of the etching process.

By applying the same approach to GaN structures where the doped layer thicknesses vary through the sample structure, we have shown that the simulation can predict chronoamperometry data for structures beyond DBRs. We have hence also suggested that a modified version of the simulation may be used to represent an alternative approach to electrochemical etching where lithographically-defined trenches are used to allow etchant access to the doped layers, and the surface is protected. The output of the modified simulation suggests that even where the lithographically-defined trench approach is used, TD-mediated etching may still be relevant and control the pore morphology. This suggests broad applicability of the simulation.

## VI. SUPPLEMENTARY MATERIAL

Figure S1 shows chronoamperometry traces recorded for the ECE of three different samples at 9 V: a standard sample with no coating; a sample whose sides were covered with polymer glue; and a sample whose sides and back were covered with polymer glue.

**DATA AVAILABILITY STATEMENT**

**Code Availability** The source code for the stochastic simulations used in this work is archived on Zenodo at https://doi.org/10.5281/zenodo.18078606. The live development version is currently maintained at https://github.com/piotr-sokolinski/Etching-GaN.

**ACKNOWLEDGEMENTS**

This research was supported by the Royal Academy of Engineering under the Chairs in Emerging Technologies Scheme, which is sponsored by the Department for Science, Innovation and Technology (DSIT). Funding was also received from the EPSRC under EP/X015300/1. BT would like to acknowledge the Ernest Oppenheimer Trust at the University of Cambridge. We acknowledge the Royce institute for the use of the Zeiss Crossbeam 540 under grant EP/R008779/1.